\documentclass[prl,aps,twocolumn,showpacs]{revtex4}

\usepackage{pxfonts}
\usepackage{graphicx}
\usepackage{color}
\usepackage{times}
\begin{document}

\title{Polariton Condensate Transistor Switch}

\author{T.~Gao$^{1,2}$, P.~S.~Eldridge$^{2}$, T.~C.~H.~Liew$^{3}$, S.~I.~Tsintzos$^{2,4}$, G.~Stavrinidis$^{2}$, G.~Deligeorgis$^{5}$, Z.~Hatzopoulos$^{2,6}$, and P.~G.~Savvidis$^{1,2}$}

\affiliation{$^{1}$Department of Materials Science \& Technology, University of Crete, Greece\\ $^{2}$IESL-FORTH, P.O. Box 1527, 71110 Heraklion, Crete, Greece\\ $^{3}$School of Physical and Mathematical Sciences, Nanyang Technological University, 637371, Singapore\\ $^{4}$Cavendish Laboratory, University of Cambridge, Cambridge CB3 0HE, United Kingdom\\ $^{5}$Universite de Toulouse, UPS, INSA, INP, ISAE, UT1, UTM and CNRS-LAAS, F-31077 Toulouse, Cedex 4, France\\ $^{6}$Department of Physics, University of Crete, 71003 Heraklion, Crete, Greece}

\begin{abstract}

A polariton condensate transistor switch is realized through optical excitation of a microcavity ridge with two beams. The ballistically ejected polaritons from a condensate formed at the source are gated using the 20 times weaker second beam to switch on and off the flux of polaritons. In the absence of the gate beam the small built-in detuning creates potential landscape in which ejected polaritons are channelled toward the end of the ridge where they condense. The low loss photon-like propagation combined with strong nonlinearities associated with their excitonic component makes polariton based transistors particularly attractive for the implementation of all-optical integrated circuits.

\end{abstract}

 \maketitle

Contemporary electronics face ever increasing obstacles in achieving higher speeds of operation. Down-scaling which has served Moore's law for decades is approaching the inherent limits of semiconductor materials \cite{Frank2001,Doering2001,Keyes1977,anis2002}. Even though a number of novel approaches \cite{Hisamoto2002,Gusev2001,Yan2011} have managed to improve operating frequency and power consumption, it is commonly acknowledged that in the future, charged carriers will have to be replaced by information carriers that do not suffer from scattering, capacitance and resistivity effects. Although photonic circuits have been proposed, a viable optical analogue to an electronic transistor has yet to be identified as switching and operating powers of these devices are typically high \cite{miller}.

Polaritons which are hybrid states of light and electronic excitations offer an attractive solution as they are a natural bridge between these two systems. Their excitonic component allows them to interact strongly giving rise to the nonlinear functionality enjoyed by electrons. On the other hand, their photonic component restricts their dephasing allowing them to carry information with minimal data loss. Notably from the view of solid state physics polaritons are bosonic particles with a particularly light effective mass. These properties allow for the condensation of polaritons into a massively occupied single low-energy state, which shows many similarities to atomic Bose Einstein condensates \cite{kasprzak, lagoudakis, amo_nature, Baumidis}. The macroscopic quantum properties of polariton condensates combined with their photonic nature make them ideal candidates for use in quantum information devices and all optical circuits \cite{liewPRL, Amo, Paraiso, Adrados, tosi}. Several recent works address the possibility of optical manipulation of polariton condensate flow however these stop short of demonstrating actual gating of polariton condensate flow a prerequisite for implementation of integrated optical circuits \cite{sanvitto,wertz, tosi,liewphysica_E}. 

In this paper, a high finesse microcavity sample fabricated into a ridge is utilized to develop an exciton-polariton condensate transistor switch. A polariton condensate formed by optical excitation serves as a source of polaritons which are ballistically ejected along the channel as shown in Figure 1(a). Polariton propagation can be controlled using a second weaker beam that gates the polariton flux by modifying the energy landscape [Fig. 1(b)]. In the absence of the gate beam the ballistically ejected polaritons can be used to efficiently feed (collector) polariton condensate forming at the edge of the ridge. Owing to the highly nonlinear nature of the (collector) condensate, attenuations approaching 90$\%$ can be achieved.

\begin{figure}[h]
\includegraphics[width=7.5cm]{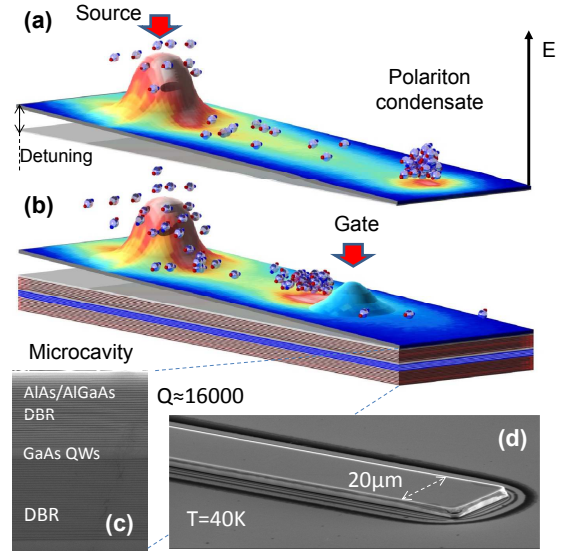}
\caption{(color online). Schematic of the polariton condensate
transistor based on a microcavity ridge (a) without and
 (b) with the gate. Scanning Electron Microscopy images, (c) the cross-section of bulk sample and (d) a 20$\mu$m ridge.}
\end{figure}

\begin{figure*}[t]
\includegraphics[width=16cm]{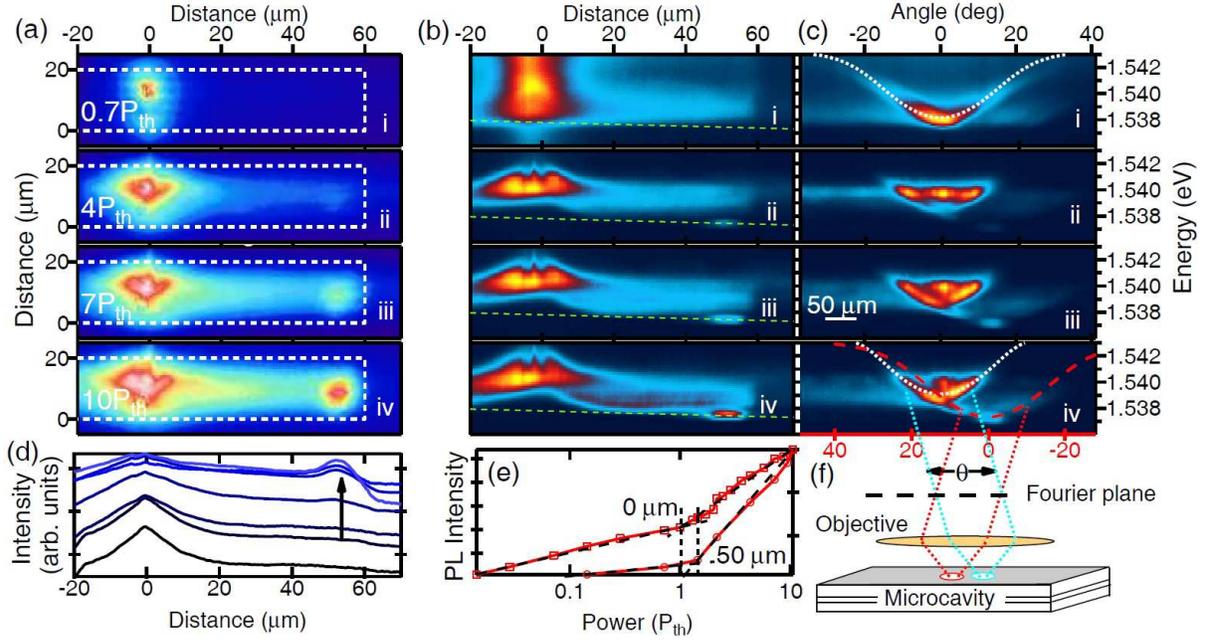}
\caption{(color online). (a) Real-space (b) energy-real space and (c) energy-angle space images for increasing source power (I-IV). Dashed/dotted lines in (c) are modelled fits to the lower polariton dispersion. (d) Normalised line profile (log scale) along the centre of the ridge for increasing source power (as indicated by the arrow). (e) The integrated intensity at the source (1st threshold 7mW) and the ridge end (2nd threshold 10mW). (f) Schematic of the experimental set up. The photonic potential in (b) corresponds to continuous variation in the detuning from -2.0\,meV at the source to -2.5\,meV at the ridge end.}
\end{figure*}

Microcavity ridges are formed through reactive ion etching of a high finesse (quality factor $>16000$) planar microcavity [Fig. 1(c),(d)], details of which can be found in \cite{tsotsis,tosi}. The ridges with dimensions 20 $\times$ 300 microns are aligned along the sample wedge so that the exciton-photon detuning produces potential gradient for polaritons. This potential gradient suppresses reflections from the end of the ridge which although produced rich physical phenomena\cite{wertz}, can cause parasitic interference if operated in the transistor regime. We note that lateral confinement in this 20\,$\mu$m wide structure is negligible compared to much thinner 1D polariton wires\cite{dasbach}. To reduce unwanted quantum well emission from the ridge sides, only the top distributed Bragg reflector (DBR) was etched and a thin layer of gold was deposited over the etched plateau. The sample is mounted in a cryostat cooled to 40K.

The source and gate beams are derived from a continuous wave Ti:Sapphire laser tuned to the first high energy Bragg mode of the DBR, focused through a microscope objective to form 2\,$\mu$m spots spatially separated by 40\,$\mu$m along the ridge. The same objective is used to collect and direct the emission toward a nitrogen cooled CCD spectrometer. Centering the ridge along the slits enables spatial and spectral analysis of the emission. Far-field imaging allows the emission to be angularly resolved for different points along the ridge. A cross-polariser in the collection path reduces scattered laser light.

We first examine propagation and condensation of polaritons along the ridge for different powers with only the source present [Fig. 2]. The source laser is placed 60\,$\mu$m from the end and the emission is resolved in real space [Fig. 2(a)], along the ridge energetically [Fig. 2(b)] and angularly [Fig. 2(c)]. The detuning shown by the dashed line in Fig. 2(b) varies linearly across the ridge to collect the ejected polaritons at the ridge end. The dotted and dashed lines for the k-space images are fits to the emission using a coupled oscillator model.

Below threshold real space image [Fig. 2(a)I] reveals the emission is concentrated at the source whilst angularly resolved emission [Fig. 2(c)I] confirms that the emission originates from the lower polariton (LP) branch. The spectrally resolved image [Fig. 2(b)I] shows the relaxation bottleneck at higher energies and evidences long polariton lifetimes through the extension of the polariton emission to the end of the ridge. Above threshold collapse of the relaxation bottleneck is observed [Fig. 2(b)II] and the dispersion displays signatures of non-equilibrium condensation as expected for a tightly focused laser spot with condensation away from k=0 [Fig. 2(c)II] \cite{wouters}.

\begin{figure*}[t]
\includegraphics[width=16cm]{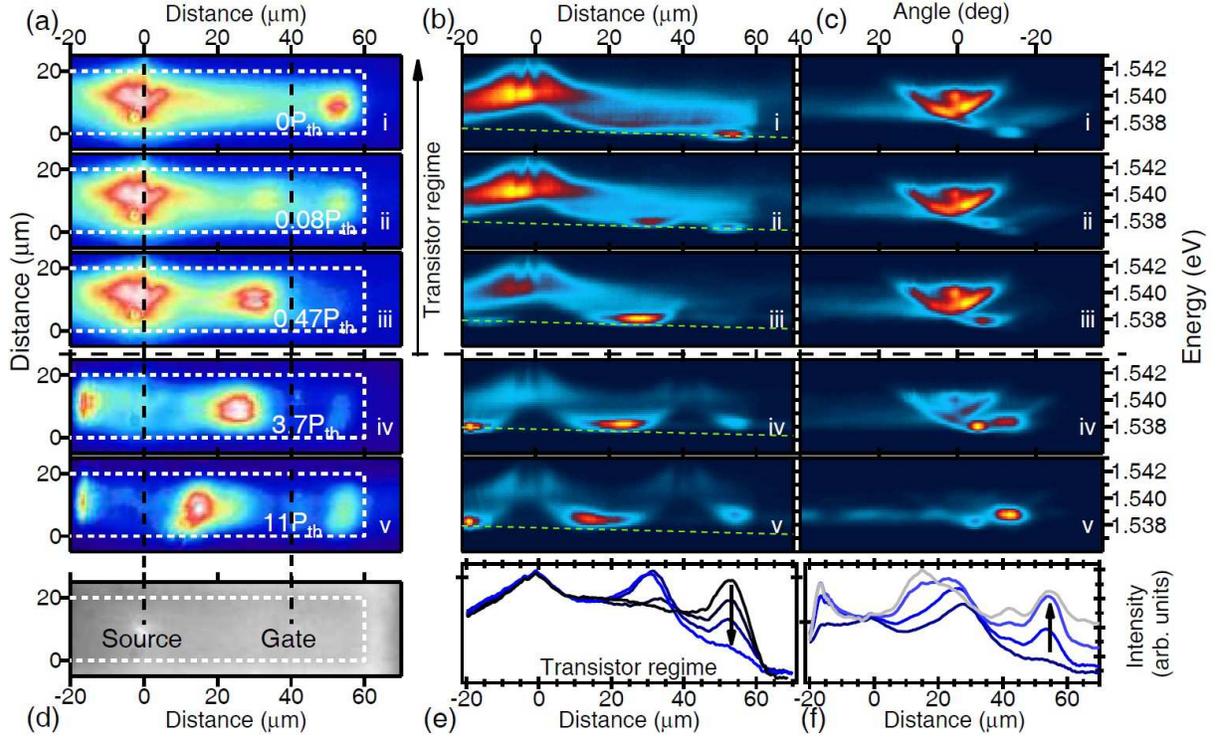}
\caption{(color online). (a) Real-space (b) energy-real space and (c) energy-angle space images for different gate powers (I-V) for a constant source power (10P$_{th}$). (d) White light image of the ridge showing the positions of the source and the gate. The line profile (log scale) along the centre of the ridge for a gate power (e) below and (f) above threshold for increasing gate power (as indicated by arrow).}
\end{figure*}

With the increasing density, the LP branch blueshift extending several microns on either side of the laser spot as seen in  Figs. 2(b)II-IV results from strong mutual repulsion of injected exciton-polaritons in the reservoir. 

This spatially localized energy blueshift ballistically ejects polaritons producing the bright constant energy feature in Fig. 2(b)II. The ejected polaritons with well defined inplane wavevector can undergo cascaded parametric scattering as in \cite{savvidis} producing bright replicas appearing at lower energies. In addition to parametric scattering several other relaxation channels such as exciton-polariton scattering \cite{porras} and phonon-polariton scattering \cite{tassone,Niccola} contribute to the relaxation of polaritons toward lower energy states efficiently feeding the polariton condensate at the end of the ridge. We have independently confirmed that at higher temperature when lasing occurs in the weak coupling regime for the same sample detuning conditions, no spatial spreading takes place confirming that above threshold the strong coupling regime is maintained throughout the sample.

The formation of the (collector) polariton condensate with increasing power can be clearly seen by extracting spatial intensity profiles plotted in Figure 2(d). The non-linear increase in the integrated intensity [Fig. 2(e)] for the emission at the ridge end and spectral narrowing (not shown) marks the onset of the second condensate. Unlike the source condensate, (collector) condensate forms at exactly the LP energy consistent with the reservoir of excitons being spatially confined to the source spot as seen in Figure 2(b)IV. This is confirmed by overlapping k-space emissions from the source and collector condensates using the experimental geometry in Figure 2(f) which allows simultaneous readout of angular and energetic content of the two emissions with an offset defined by the schematic.
As seen in Figure 2(c)IV the angular emission at each spot can be well fitted using density dependent dispersions shown with dashed and dotted lines.

Transistor operation is achieved when a weak gate beam is introduced while keeping the source power constant ($10P_{th}$) [Fig. 3(d)]. A nearly complete switching off of the polariton flux feeding the collector condensate [Fig. 3(a)I-III] is achieved by controlling the energy barrier at the gate spot via density dependent blueshift [Fig. 3(b)I-V].
Maximal switching is realized for sub-threshold gate powers ($0.5P_{th}$) before the polaritons ejected by the gate start feeding the collector condensate, at which point the transistor regime breaks down [Fig. 3 IV-V].
The gating of the polariton condensate is evidenced from the spectra [Fig. 3(e)] showing normalised line profiles for different gate powers from Figure 3(a). The condensate emission at 50$\mu$m decreases as polariton flux from the source is blocked by the gate. The blocked polaritons equilibrate at a local minima of potential landscape which drifts with the increasing gate power as seen from both real space images in Fig. 3(a),(b) and angular-resolved images in Fig. 3(c).

For the highest gate powers, the system is expected to undergo more efficient energy relaxation and in this regime, shown by Figs. 3(a)V, 3(b)V and 3(c)V, the system relaxes to the ground state. Theoretically, the ground state is described by the Gross-Pitaevskii equation for the polariton mean-field, $\psi(x)$:

$$i\hbar\frac{d\psi(x)}{dt}=(\widehat{E}+V(x)+\alpha\mid\psi(x)\mid^{2})\psi(x)$$

where $\widehat{E}$ is the kinetic energy operator of polaritons,
which reproduces the lower polariton non-parabolic dispersion
observed experimentally using a coupled mode model.
$V(x)=V_{0}(x)-\beta x+P(x)$ is the effective potential
experienced by polaritons, which is composed of: the confinement
potential $V_{0}(x)$ that represents the walls of the ridge; the
potential gradient, $-\beta$; and a pump induced blueshift
\cite{wouters, ferrier} $P(x)$, which is composed of Gaussian
spots with intensity and position corresponding to the lasers used in the experiment. Pump-induced blueshifts have previously been shown to allow the engineering of the polariton potential using resonant\cite{amo} and non-resonant\cite{ferrier} excitation. In our experiments, the pump excites high energy excitons that induce a blueshift of polaritons through interactions. The polariton-polariton interaction strength, $\alpha$, can be estimated by the formula \cite{tassonePRB99} $\alpha\approx6E_Ba_B^2$ using standard values for the exciton binding energy ($E_B$) and Bohr radius ($a_B$) in GaAs quantum wells. Given $\alpha\approx 0.004$meV$\mu m^2$, the polariton density at the collector spot can be estimated from the observed blueshift as $7.5\times10^9$cm$^{-2}$. For a fixed number of polaritons, the Gross-Pitaevskii equation can be solved numerically\cite{footnote} using imaginary time propagation to yield the ground state of the system. The results are shown in Fig. 4.

\begin{figure}[t]
\includegraphics[width=7.5cm]{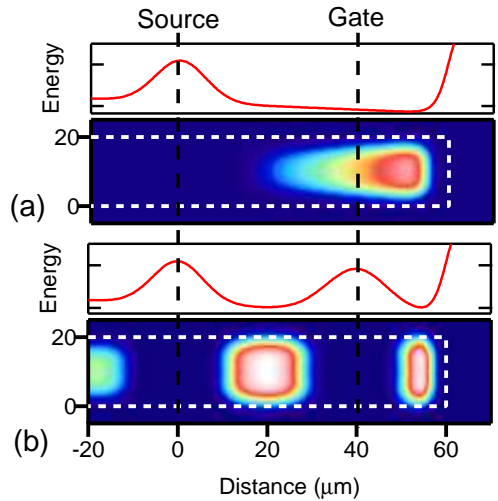}
\caption{(color online). (a) The polariton potential landscape and solutions to the Gross-Pitaevskii equation for the source beam (a) and the source and gate beams (b).}
\end{figure}

For the case of a single pump, the ground state of the polariton
condensate forms at the lower energy end of the ridge, as shown in
Fig. 4(a). This corresponds to the spot observed experimentally in
Fig. 2(a) for high pump powers. The Gross-Pitaevskii equation does
not predict the emission from the pump spot position seen in the
experiment since this component derives from non-ground state
condensation \cite{krizanovskii, margou}.

The presence of a second laser pump can strongly affect the
observed ground state, by increasing the energy locally in the
region where the condensate would otherwise form. For sufficient intensity of the second pump, one can arrive at the situation where three local minima in the potential exist: one on either side of the two pump spots and one between them [Fig. 4(b)].

Although qualitatively clear, the regime of incomplete energy relaxation is challenging to describe quantitatively, requiring a dynamical description of non-equilibrium condensation \cite{woutersPRL} in addition to energy relaxation within the condensed polariton modes. Existing techniques are based on polariton-exiton scattering \cite{woutersPRB2010} rather than polariton-phonon scattering or coherent pumping \cite{savenko}.

We now discuss the performance of our prototype polariton condensate transistor. The polariton dispersion yields a group velocity of 2.5$\mu$m/ps, nearly 30 times faster than the speed of electrons in modern silicon transistors, ensuring a short transit time from gate to collector (10ps). Therefore in the current scheme the switching time of the  device is limited by the polariton lifetime (18ps), allowing operation frequencies of tens of gigahertz, one order of magnitude faster than those of exciton based transistors\cite{high}. Further work is in progress toward reduction of operation powers of the polariton transistor through either resonant condensate injection or implementation of electrically controlled gating. 

In summary, a prototype polariton condensate transistor is realized utilizing a high finesse microcavity sample fabricated into ridges. We show that the polariton flux from a condensate can be controlled using a gate beam that is 20 times weaker than the source. This spatial control with attenuation approaching 90$\%$ allows the implementation of all-optical polariton transistor suitable in both optoelectronic and all-optical schemes \cite{liewPRL, liewPRB}. In the short-term they could be used in optical inter-core communication for transferring data within and between chips. In the long-term they could enable the elimination of electronics all together.

We note that a different polariton transistor scheme has also been recently demonstrated through resonant excitation of a planar microcavity in the optical discriminator regime \cite{Ballarini}.

We acknowledge funding from the FP7 Clermont 4 and INDEX ITN projects and the POLATOM ESF Research Network Programme. P.S would like to thank M. Kaliteevski for fruitful discussions and acknowledges funding from EU FP7 IRSES POLATER grant. This work has also been co-funded by the EU and National Resources through scientific program "EPEAEK II-"HRAKLEITOS II, University of Crete".


\begin{thebibliography}{}
\bibitem{footnote} The maximum pump-induced (reservoir-induced) blueshift was taken as 2meV, estimated from the observed blueshift at the pump position. $\beta$=9\,meV/mm.
\end{thebibliography}
\end{document}